# HR-EBSD analysis of in situ stable crack growth at the micron scale

Abdalrhaman Koko[a*], Thorsten H. Becker[b], Elsiddig Elmukashfi[c], Nicola M. Pugno[d,e], Angus J. Wilkinson[a], and T. James Marrow[a]

[a]  Department of Materials, University of Oxford, Oxford OX1 3PH, United Kingdom

[b]  Centre for Materials Engineering, Department of Mechanical Engineering, University of Cape Town, Cape Town, South Africa

[c]  Department of Engineering Science, University of Oxford, Oxford OX1 3PJ, United Kingdom

[d]  Laboratory for Bioinspired, Bionic, Nano, Meta Materials & Mechanics, Department of Civil, Environmental and Mechanical Engineering, University of Trento, Via Mesiano 77, 38123 Trento, Italy

[e]  School of Engineering and Materials Science, Queen Mary University of London, London E1 4NS, United Kingdom

## Abstract

Understanding the local fracture resistance of microstructural features, such as brittle inclusions, coatings, and interfaces, at the microscale is critical for microstructure-informed design of materials. In this study, a novel approach has been formulated to decompose the $J$-integral evaluation of the elastic energy release rate to the three-dimensional stress intensity factors directly from experimental measurements of the elastic deformation gradient tensors of the crack field by in situ high (angular) resolution electron backscatter diffraction (HR-EBSD). An exemplar study is presented of a quasi-static crack, inclined to the observed surface, propagating on low index $\{hkl\}$ planes in a (001) single crystal silicon wafer.

---

[*] Corresponding author. E-mail address: abdo.koko@materials.ox.ac.uk

## 1. Introduction

Since the earliest work of Griffith [1], it has been known that a sharp crack can propagate when the available global energy is sufficient to provide the necessary surface and dissipated energy (i.e., thermodynamic energy balance). This criterion is described by the critical strain energy release rate, $G_c$, and in linear elastic materials, it is represented by the critical stress intensity factor (SIF), $K_{IC}$ [2]. Non-local energy-based methods – based on the Noether theorem – subsequently evolved this isotropic linear elastic fracture mechanics to address anisotropic nonlinear problems. Noether [3] proved that within a Euclidean homogenous and isotropic reference frame (i.e., Noether frame) where energy and momentum are conserved, a path-independent integral describes the potential energy stored by the force field (i.e., an equivalent descriptor of forces). A mathematical formulation of the conservation laws, which originated from Günther [4], was applied to elastostatics in the form of path-independent integrals of some functionals of the elastic field over the bounding surface of a closed region. Some years earlier, Eshelby [5,6] represented the 'force on an elastic singularity or inhomogeneity' in the form of a surface integral that, in the absence of other defects, lead to an elastostatic conservation law. The two-dimensional analogue of this conservation law was introduced by James Rice [7] as a path-independent line integral that could be applied to a stress concentration. However, earlier investigations by Sanders [8] and Cherepanov [9] were closely related. Rice's integral has, subsequently, become known as the $J$-integral and is widely adopted in fracture mechanics.

Hutchinson [10] and Rice and Rosengren [11] were the first to apply the $J$-integral as a criterion to describe the strain energy release rate for crack growth in linear elastic or elastic-plastic materials. Its application was limited to loaded cracks with no internal stress/strains or edge tractions [12] and no significant plasticity [13], as these are the necessary conditions for the integral to be path-independent. Nevertheless, with modifications, the $J$-integral can be applied to other conditions, including significant plasticity [14]. The numerical character of the $J$-integral facilitates its evaluation with finite element (FE) methods, typically by application of the divergence theorem and equivalent domain integration (EDI) [15].

Application of the $J$-integral to mixed-mode crack problems (e.g., where crack kinking or deflection occurs, and the crack tip is not loaded purely by opening) is complex since the $J$- integral does not distinguish between the strain energy contributions from crack opening (mode I) and shearing (modes II and III). Thus, decomposition is needed to evaluate the two- or three- dimensional stress intensity factors that describe the equivalent elastic field. In the early 1970s, Bueckner [16] and Rice [17] used a weight function concept for 3D stress analysis of anisotropic linear elastic materials under combined mode I and II loading. Further investigation by Cottrell and Rice [16] showed that the total elastic strain energy release rate is the sum of the energy release rates of the independent modes in mixed mode loading, if the crack tip retains the same configuration (i.e., no change in direction) and there is small-scale yielding. Shih and Asaro [17] introduced the interaction integral method to separate the deformation fields by superimposing an auxiliary field onto the actual tensor field, thus decomposing the field in terms of its symmetric (mode I), in-plane antisymmetric (mode II), and out-of-plane antisymmetric (mode III) parts.

The critical $J$-integral ($J_{IC}$) – at the condition of quasi-static crack propagation – is widely used as a fracture criterion for different materials [18,19] and test configurations [20,21], where it is coupled with analytical or finite element analyses that use knowledge of remote applied boundary conditions and specimen geometry to perform structural integrity assessments. Recently, methods have been developed that use experimental data for the displacement fields around cracks as local boundary conditions in finite-element models to solve the elastic strain field and calculate the $J$-integral. These analyses demonstrated that the potential elastic strain energy release rate could be quantified by such local measurements without knowledge of the external boundary conditions (i.e., load, crack length) [22,23]. Such local analyses are valuable when the external conditions are unknown or uncertain.

This local analysis has also been extended to use experimental data for elastic strain fields that were mapped by synchrotron X-ray diffraction [24,25]; the elastic strains were integrated to find the equivalent elastic displacement field that provides the missing derivatives (i.e., $u_{2,1}$) required for $J$-integral analysis [26]. The diffraction strain mapping in these studies was done at a relatively large scale (cm-size specimens with mm-size cracks) and with low spatial resolution. To study the criteria for crack growth at the micron scale within the microstructure, for instance, in local investigations of the toughness of brittle phases or coatings, we need high-resolution data. It is also necessary to characterise the stress and strain fields in situ at, or approaching, the critical state for crack propagation. Furthermore, the analysis needs to be appropriate to inclined crack planes that are not necessarily well oriented for surface observations. Such an approach could replace or complement existing micro-mechanical test methods that rely on analytical solutions; these require knowledge of the applied loads and displacements, sample geometry, crack length, etc., [27,28] that can be difficult to obtain accurately. Micro-mechanical tests, which are primarily focused on mode I loading, may also be dependent on size and affected by focused ion beam (FIB) -milling damage [28,29], and can be influenced by non-symmetrical loading [30] and user-bias [31].

High-resolution electron backscatter diffraction (HR-EBSD) is used to study deformation states in crystalline materials. It is a non-destructive surface analysis method that can quantify elastic strains and lattice rotations at a sensitivity of $\pm 10^{-4}$ [32,33] with high-resolution mapping of cross-correlated electron backscatter diffraction patterns (EBSPs). Recent studies have shown that the $J$-integral can be used to parameterise (in situ) the two-dimensional strain fields, measured by HR-EBSD, of the singularities of deformation twins [34] and sip bands [35]. However, these analyses did not consider the three-dimensional nature of the stress concentrator, such as its inclined plane or the effect of out-of-surface displacements on the near-surface strain field.

This study presents a novel approach to calculating and decomposing the $J$-integral elastic strain energy release rate to the three-dimensional stress intensity factors directly from the elastic deformation gradient tensors obtained from the near-surface membrane by HR-EBSD. As a demonstration, we parametrise the fields ahead of a crack, observed in situ while propagating under quasi-static, mixed-mode conditions on an inclined plane within a (001) silicon single crystal wafer.

## 2. Method

### 2.1. Materials and Experimental Details

The exemplar sample (Figure 3 and Table 1), measuring 9.7 x 8.9 x 0.5 mm$^3$, was cleaved from a single crystal silicon wafer with a thickness of 0.5 mm. Its pre-polished surface was parallel to $(001)$ (Figure 1a). The loaded edges of the sample, parallel to $[110]$, were abraded manually. The sample was positioned between the jaws of a 2 kN Deben® 70° pre-tilted loading stage inside a Carl Zeiss Merlin field emission gun scanning electron microscope (FEG-SEM). The SEM chamber and loading stage were plasma cleaned and purged for 1 hour before the sample was loaded by compression in the $[\bar{1}10]$ direction in displacement-control by the movement of one jaw. The loading speed, initially 0.1 mm min$^{-1}$, was increased to 0.2 mm min$^{-1}$ after some initial damage occurred (the red line in Figure 1b indicates the change). Once a stable propagating crack was established, the loading was arrested at fixed displacements to allow the acquisition of electron backscatter patterns (EBSPs) to map the vicinity of the crack tip whilst under load.

The EBSPs were collected using a Bruker eFlash CCD camera using conditions of a 20 kV/10 nA beam, 18 mm working distance, and 200 milli-second exposure time per pattern. The step size was 0.25 μm, so the 800 x 600-pixel area mapped was about 20 x 15 μm$^2$. Load relaxation during the EBSD mapping was negligible (<2.8%). For the exemplar sample, the crack was propagated incrementally in 12 intervals (indicated by the dotted blue lines in Figure 1b). No correction was done for the frame compliance; thus, the plotted uniaxial stress and strain should be taken only as an indicator of stress increment.

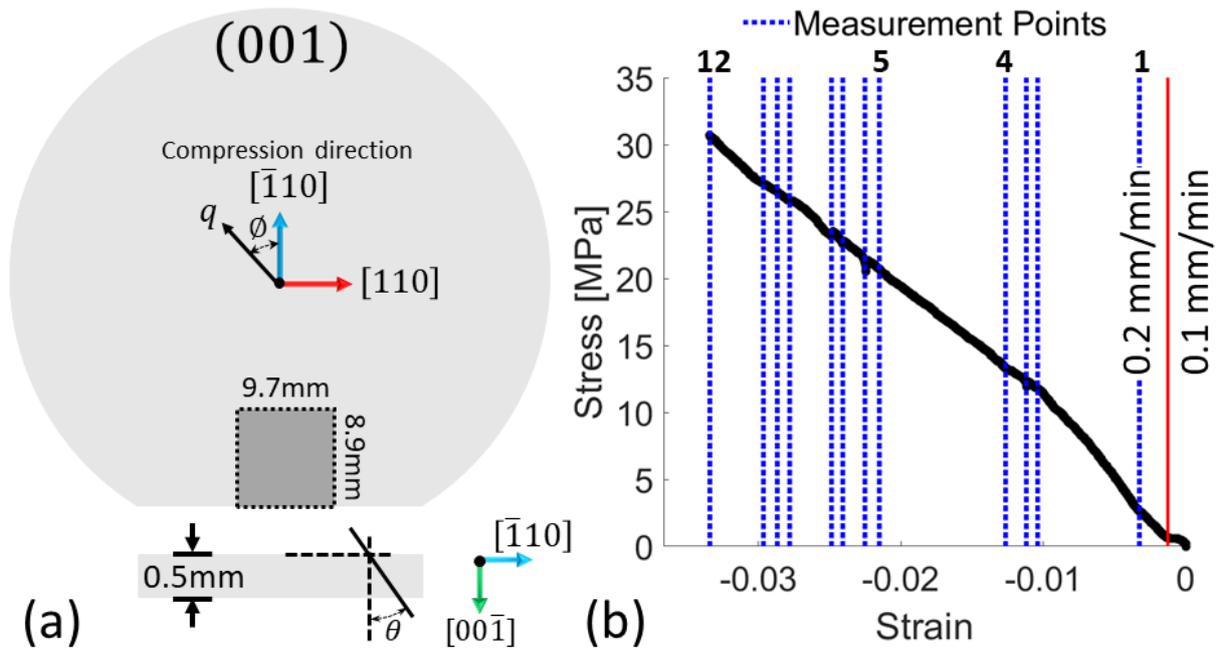

*Figure 1: (a) Schematic of pre-polished silicon (001) wafer with the 9.7 x 8.9 x 0.5 mm³ cleaved specimen that was loaded in compression parallel to [Ī10]. The crack trace angles, ϕ, measured with respect to the x-axis and θ, measured with respect to the z-axis are shown. (b) Nominal uniaxial stress and strain (crosshead displacement/sample dimension). Loading started at a low speed of 0.1 mm min⁻¹, and the minor load dip near the red line in Figure 1b is due to the detachment of a small fragment at the loading contact. The speed was then increased to 0.2 mm min⁻¹ (at the red line in b). The blue lines mark the collection of EBSD data when loading was suspended at fixed displacement.*

Focused Ion Beam (FIB) milling was used to investigate the sub-surface orientation of the crack. This was done using a Zeiss Auriga dual-beam SEM-FIB system that has a Schottky field emission Gemini electron column coupled with an Orsay Physics "Cobra" Ga+ ion FIB. The sample was tilted to 54° and positioned at a working distance of 5 mm. At each point of investigation, platinum and carbon protective layers were deposited with individual thicknesses of ~1.5 µm (240 pA/ 30 kV beam). A 35 µm deep trench was cut by rough (16 nA/ 30 kV) and then fine milling (600 pA/ 30 kV). The crack trace on the trench wall, which is parallel to [00Ī] was measured with in-lens secondary electron (SE) imaging conditions at 36° tilt correction (i.e., an effective 90° view) over a depth of 15 µm. All angles and distances were measured using ImageJ [36], and the uncertainties in angle and crack increment measurement were less than ±0.5° and ±0.5 µm, respectively.

The elastic deformation field was calculated by iterative cross-correlation of the EBSPs with a reference pattern (EBSP₀) [33,37], acquired remotely from the crack and assumed to be stress-free [38] and minimising errors from pattern centre (PC) shift due to beam movement during acquisition [39]. The interplanar spacing and zone axes changes were measured in 30 independent regions of interest in each EBSP, with bicubic interpolation to obtain the best-fit solution. The difference from the reference EBSP$_0$ was related to the elastic deformation gradient tensor ($F_{ij}$), which was polarly decomposed with an average sensitivity of $1.6 \times 10^{-4}$ to deviatoric strains, $\varepsilon_{ij}$, (symmetric part, where $ij = ji$) and lattice rotations $\omega_{ij}$ (asymmetric part, where $ii = jj = 0$) [32,33]. Conditions of a traction-free surface ($\sigma_{33} = 0$) was assumed [40], and anisotropic elastic properties were obtained using (001) silicon orthotropic stiffness tensor ($C_{11} = 165.7, C_{44} = 79.6, C_{12} = 63.9$ in GPa [41]). The full deviatoric-strain and elastic stress components represent the surface membrane sampled by the backscattered electrons. The geometrically necessary dislocation (GND) distribution was estimated as detailed in [42].

## 2.2. Numerical Evaluation of the $J$-Integral

To simplify the analysis of the $J$-integral energy release rate, the elastic deformation gradient tensor, $F_{ij}$, was transformed to resolve the field to axes relative to the plane of the crack that was defined by its trace on the FIB section and its propagation direction, $q_1$; the latter obtained from its trace on the pre-polished surface. The transformed elastic deformation gradient tensor, $F'_{ij}$, was obtained by $Q_z$ transformation and then $Q_x$ transformation of equations (3) to (5), thus, $x_3 \perp q_1 \parallel x_1$, where $\theta$ is the angle between the crack plane and the $z$-axis and $\emptyset$ is the angle between the surface crack trace and the $x$ axis (Figure 1a). For convenience, the superscript $F'_{ij}$ will be omitted in the subsequent discussion, such that $F_{ij}$ refers to the elastic deformation gradient tensor after transformation to the frame of reference of the crack. The same transformations were also applied to the stiffness tensor [43].

$$Q_{z(\emptyset)} = \begin{bmatrix} \cos\emptyset & \sin\emptyset & 0 \\ -\sin\emptyset & \cos\emptyset & 0 \\ 0 & 0 & 1 \end{bmatrix} \qquad (1)$$

$$Q_{x(\theta)} = \begin{bmatrix} 1 & 0 & 0 \\ 0 & \cos\theta & \sin\theta \\ 0 & -\sin\theta & \cos\theta \end{bmatrix} \qquad (2)$$

$$F'_{ij} = Q_{pi}Q_{qj}\,F_{ij} \tag{3}$$

Evaluating the $J$-integral required computation of the $q_1$-field and formulation of the equivalent domain integration (EDI) with the domain expanding from the crack tip. The $q_1$ vector undergoes a smooth linear spatial variation across the domain ($\frac{dq_1}{dx_1}$ in (4)), with magnitude unity inside and zero outside the domain. The virtual crack propagation direction, $q_1$, is assumed to be in the positive $x_1$ direction and the crack tip is located centrally in a square contour of the mapped strain field, defined by elements (Figure 2) [44]. These conditions simplify the EDI $J$-integral analysis to equation (4) using a regularised mesh of zero thickness (i.e., surface measurement assuming plane stress conditions) where no re-mapping to the element nodes is required.

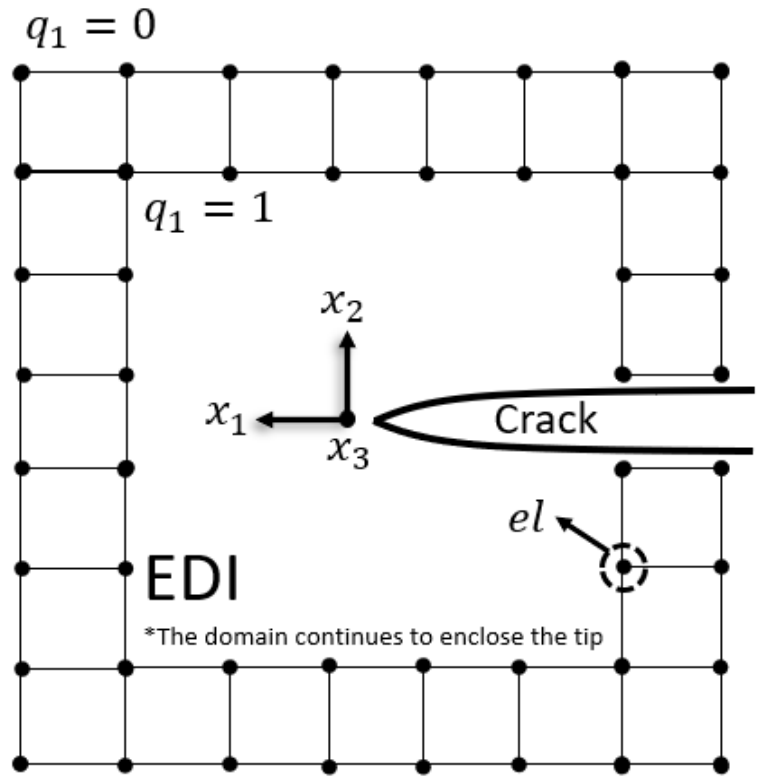

Figure 2: Schematic of the crack and the equivalent domain integral (EDI) method to calculate the potential elastic strain energy release rate or the $J_1$ or $J$-integral (so-named "J" for James Rice) for a monotonically loaded crack extending in $q_1$ direction as shown in equation (4), where $\sigma_{ij}$ is the stress tensors, $W$ is the elastic strain energy density, and $u_{i,j}$ is displacement/distortion gradient. The area increment along the equivalent domain is $dA$ and each element node el is at a point on the EBSD acquisition grid (step size 0.25 µm). The axes are defined relative to the crack propagation direction.

$$J = \sum_{el=1}^{N_{el}} \left[ (\sigma_{11} u_{1,1} + \sigma_{12} u_{2,1} + \sigma_{13} u_{3,1} - W) \frac{dq_1}{dx_1} \right.$$

$$\left. + (\sigma_{22} u_{2,1} + \sigma_{12} u_{1,1} + \sigma_{23} u_{3,1}) \frac{dq_1}{dx_2} \right] dA \tag{4}$$

As noted in the introduction, in linear-elastic materials (or with small-scale yielding conditions), the $J$-integral can be related to the three-dimensional elastic stress intensity factors using a mode-decomposition technique that separates the elastic field to symmetric (mode I), in-plane skew-symmetric (mode II), and out-of-plane skew-symmetric (mode III) components. This is done by superimposing an auxiliary field (i.e., $\bar{u}$, $\bar{F}$, and $\bar{\varepsilon}$ in equation (8), (9) and (11), relatively) onto the total field, mirrored along the $q_1$ vector (parallel to $x_1$-axis) with mesh symmetry due to the equally spaced grid. The stress intensity factors are directly related to the appropriate mode-specific $J$-integrals (i.e., $J^I$, $J^{II}$, and $J^{III}$) as in equation (5), using values of Youngs' modulus ($E$), shear modulus ($\mu$) and Poisson's ratio ($\nu$) estimated from the anisotropic stiffness tensor after being transformed to the crack plane normal [45]. The effective stress intensity factor $K_{\text{eff}}$ is directly calculated from the total energy release rate ($J$) as in equation (6) assuming plane stress conditions.

$$J = J^I + J^{II} + J^{III} = \frac{K_I^2}{E} + \frac{K_{II}^2}{E} + \frac{K_{III}^2}{2\mu} \tag{5}$$

$$K_{\text{eff}} = \sqrt{JE} \tag{6}$$

We decompose the deformation gradient tensor obtained from HR-EBSD to solve equation (4) by following a similar derivation applied to displacement and stress fields [46–49]. The displacement derivatives ($u_{i,j}$) were expressed using $F_{ij}$ and the Kronecker delta ($\delta$) (which equals unity when $i = j$) as in equation (7). These derivatives were split into modes I, II and III. The mode decomposition of normal out-of-plane deformation components is debatable, where Nikishkov and Atluri [46], Shivakumar and Raju [47], Huber *et al.* [48], Rigby and Aliabadi [49] related the asymmetrical portion to mode III; Rigby and Aliabadi [50] showed that for mode III the strain $\varepsilon_{33}^{III}$ and stress $\sigma_{33}^{III}$ are both equal to zero at the crack front. We therefore adopt this when decomposing $F_{33}$, and equation (9) gives the displacement derivatives required for equation (4).

$$u_{i,j} = u_{i,j}^I + u_{i,j}^{II} + u_{i,j}^{III} = \sum_M F_{ij}^M - \delta_{ij}^M, \quad i,j = 1,2,3, \qquad M = I, II, III \tag{7}$$

$$u_{i,j} = \frac{1}{2}\begin{pmatrix} u_{1,1} + \bar{u}_{1,1} & u_{1,2} + \bar{u}_{1,2} & u_{1,3} + \bar{u}_{1,3} \\ u_{2,1} - \bar{u}_{2,1} & u_{2,2} - \bar{u}_{2,2} & u_{2,3} - \bar{u}_{2,3} \\ u_{3,1} + \bar{u}_{3,1} & u_{3,2} + \bar{u}_{3,2} & u_{3,3} + \bar{u}_{3,3} \end{pmatrix}$$

$$+ \frac{1}{2}\begin{pmatrix} u_{1,1} - \bar{u}_{1,1} & u_{1,2} - \bar{u}_{1,2} & 0 \\ u_{2,1} + \bar{u}_{2,1} & u_{2,2} + \bar{u}_{2,2} & 0 \\ 0 & 0 & u_{3,3} - \bar{u}_{3,3} \end{pmatrix} \tag{8}$$

$$+ \frac{1}{2}\begin{pmatrix} 0 & 0 & u_{1,3} - \bar{u}_{1,3} \\ 0 & 0 & u_{2,3} + \bar{u}_{2,3} \\ u_{3,1} - \bar{u}_{3,1} & u_{3,2} - \bar{u}_{3,2} & 0 \end{pmatrix}$$

$$u_{i,j} = \frac{1}{2}\begin{pmatrix} (F_{11} + \bar{F}_{11}) - 1 & F_{12} + \bar{F}_{12} & F_{13} + \bar{F}_{13} \\ F_{21} - \bar{F}_{21} & F_{22} - \bar{F}_{22} & F_{23} - \bar{F}_{23} \\ F_{31} + \bar{F}_{31} & F_{32} + \bar{F}_{32} & (F_{33} + \bar{F}_{33}) - 1 \end{pmatrix}$$

$$+ \frac{1}{2}\begin{pmatrix} F_{11} - \bar{F}_{11} & F_{12} - \bar{F}_{12} & 0 \\ F_{21} + \bar{F}_{21} & (F_{22} + \bar{F}_{22}) - 1 & 0 \\ 0 & 0 & F_{33} - \bar{F}_{33} \end{pmatrix} \tag{9}$$

$$+ \frac{1}{2}\begin{pmatrix} 0 & 0 & F_{13} - \bar{F}_{13} \\ 0 & 0 & F_{23} + \bar{F}_{23} \\ F_{31} - \bar{F}_{31} & F_{32} - \bar{F}_{32} & 0 \end{pmatrix}$$

Finally, to calculate the strain energy density, W, the Green-Lagrangian strain tensor is split from the deformation (or displacement) gradient tensors as in equations (10) and (11).

$$\varepsilon_{ij} = \frac{1}{2}\left(F_{is}^T F_{sj} - \delta_{ij}\right) \approx \frac{1}{2}\left(u_{i,j} + u_{j,i}\right) \tag{10}$$

$$\varepsilon_{ij} = \varepsilon_{ij}^I + \varepsilon_{ij}^{II} + \varepsilon_{ij}^{III}$$

$$\varepsilon_{ij} = \frac{1}{4}\begin{Bmatrix} \varepsilon_{11} + \bar{\varepsilon}_{11} & \varepsilon_{12} - \bar{\varepsilon}_{12} & \varepsilon_{13} + \bar{\varepsilon}_{13} \\ \varepsilon_{21} - \bar{\varepsilon}_{21} & \varepsilon_{22} + \bar{\varepsilon}_{22} & \varepsilon_{23} - \bar{\varepsilon}_{23} \\ \varepsilon_{31} + \bar{\varepsilon}_{31} & \varepsilon_{32} - \bar{\varepsilon}_{32} & \varepsilon_{33} + \bar{\varepsilon}_{33} \end{Bmatrix} + \frac{1}{4}\begin{Bmatrix} \varepsilon_{11} - \bar{\varepsilon}_{11} & \varepsilon_{12} + \bar{\varepsilon}_{12} & 0 \\ \varepsilon_{21} + \bar{\varepsilon}_{21} & \varepsilon_{22} - \bar{\varepsilon}_{22} & 0 \\ 0 & 0 & \varepsilon_{33} - \bar{\varepsilon}_{33} \end{Bmatrix}$$

$$+ \frac{1}{4}\begin{Bmatrix} 0 & 0 & \varepsilon_{13} - \bar{\varepsilon}_{13} \\ 0 & 0 & \varepsilon_{23} + \bar{\varepsilon}_{23} \\ \varepsilon_{31} - \bar{\varepsilon}_{31} & \varepsilon_{32} + \bar{\varepsilon}_{32} & 0 \end{Bmatrix} \tag{11}$$

The mode-specific full deviatoric-strain tensors ($\varepsilon_{ij}^M$) and elastic stress ($\sigma_{ij}^M$) are then related by Hooke's law with the anisotropic stiffness matrix.

## 3. Results

The exemplar crack was propagated under quasi-static conditions by in-plane compression of a (001) single crystal silicon wafer (see method). Figure 3a shows the fracture surface and the locations of the in situ HR-EBSD measurements and subsequent FIB trench cuts that revealed its sub-surface inclination. The nominal applied stress, and the orientation and crack length increment at each observation are summarised in Table 1. After some initial changes, the crack's orientation was practically constant over ~700 µm (labels 5 to 12), with the propagation direction within 1° of $[\bar{3}10]$ and the crack plane within 4° of (131).

After breaking open the sample, inspection of the fracture surface (Figure 3) shows that the crack plane was generally smooth over the propagation region of interest following the complex initiation of fracture. Small steps (beach marks similar to 'Wallner' lines) indicate the crack front was essentially perpendicular to the EBSD-observed (001) surface. Wing cracks can be observed, particularly between locations 5 to 9 (Figure 3b).

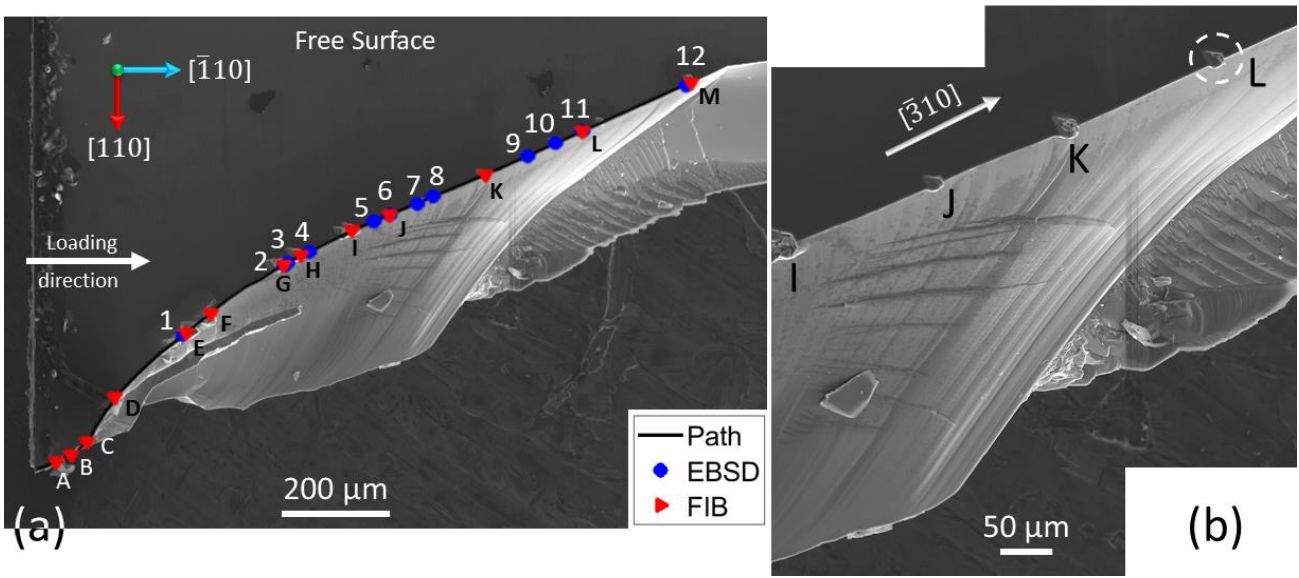

*Figure 3: SEM images of the fracture surface of the exemplar crack, which was propagated in a (001) silicon wafer loaded in compression parallel to $[\bar{1}10]$: a) showing the fracture surface and HR-EBSD mapping and FIB cutting locations after fracture. The numbers mark the intervals of crack growth, after which the stable crack tip was observed under load by HR-EBSD. The sample surface was cleaned from the accumulated debris before FIB cutting of trenches parallel to (110) at the marked locations to examine the sub-surface crack geometry, which shows details of the surface topography (beach marks and river lines). The locations of the FIB-milled sites are labelled, and the FIB site (L) is circled in (b).*

The example of an in situ HR-EBSD observation of the crack (Figure 4) shows the elastic plane stress components (Figure 4a) and the decomposed mode I, II, and III von Mises stresses after rotational transformations (Figure 4b). Here, the field has been rotated ($Q_z$) to align the crack trace with the $x_1$-axis and translated to place the crack tip at the centre of the map, which has been cropped to 15 x 15 µm². The same stress field, after rotation ($Q_x$) to align $x_3$ perpendicular to the crack plane, is shown in Figure 4c. The most significant effect is observed in the shear components; the magnitude of in-plane shear ($\sigma_{xy}$) has increased and the out-of-plane shear ($\sigma_{xz}$) has decreased. The mode II von Mises stress is higher than the von Mises stresses of mode I and III.

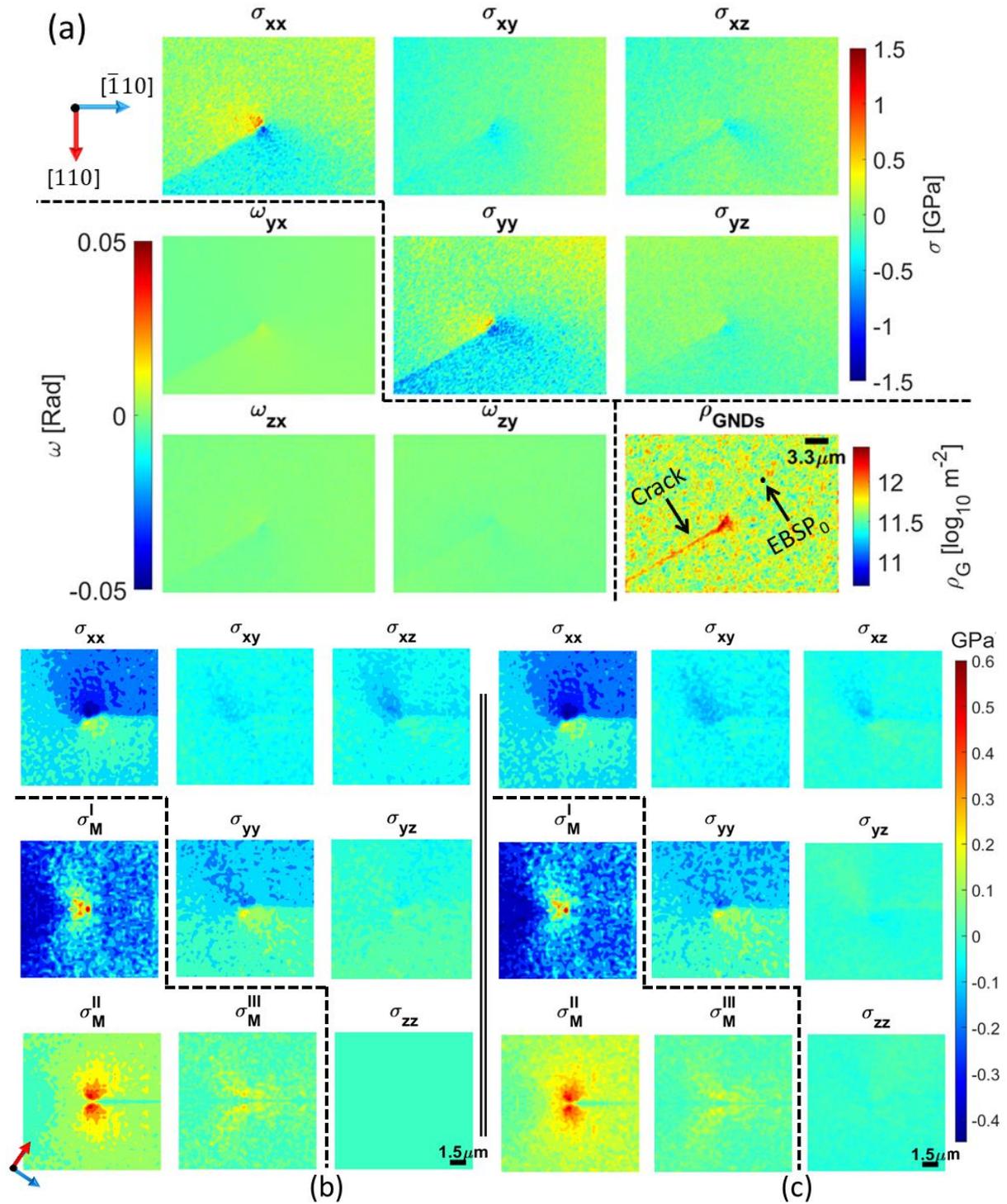

*Figure 4: Example HR-EBSD data (Figure 3a, at label 1 - see Table 1): a) elastic stress components ($\sigma_{ij}$), rotation ($\omega_{ij}$) and geometrically necessary dislocation (GND) density ($\rho_{GNDs}$) calculated in the microscope xyz reference frame, showing high normal stresses in the surface plane. The stress normal to the surface is zero. The average GND density is low and is high only along the crack edges and at its tip; b) elastic stress components and decomposed mode I, II, III von Mises stresses ($\sigma_M^{I,II,III}$) after rotation by $\emptyset = 44.6°$; and c) then by $\theta = 20.6°$ (defined in Figure 1). The crack propagation is from left to right, compared to the reference frame data presented in (a). The rotations, coupled with regularised pattern acquisition (square meshing where $dA$ equals the step size), and centred crack tip ease the implementation of the domain equivalent integral (EDI) method of J-integral analysis.*

$J$-integral values and the three-dimensional elastic stress intensity factors were calculated from each EBSD-measured field as the integration domain was advanced from the crack tip (Figure 5). The initial values with the smallest domain are erroneous due to noise and large strains close to the crack, and convergence is obtained as the domain expands. Contours larger than 2 μm correspond to encapsulation of the highly deformed field at the tip, and these stable values were used. The mode-specific $J$-integral (or equivalent stress intensity) is the average of the 10 values in the converged shaded region (Figure 5). The total $J$-integral is the summation of the three mode-specific $J$-integrals (see method). The average values (and standard deviation) for each of the observations (labelled in Figure 3) of the quasi-static propagating cleavage crack are given in Table 1.

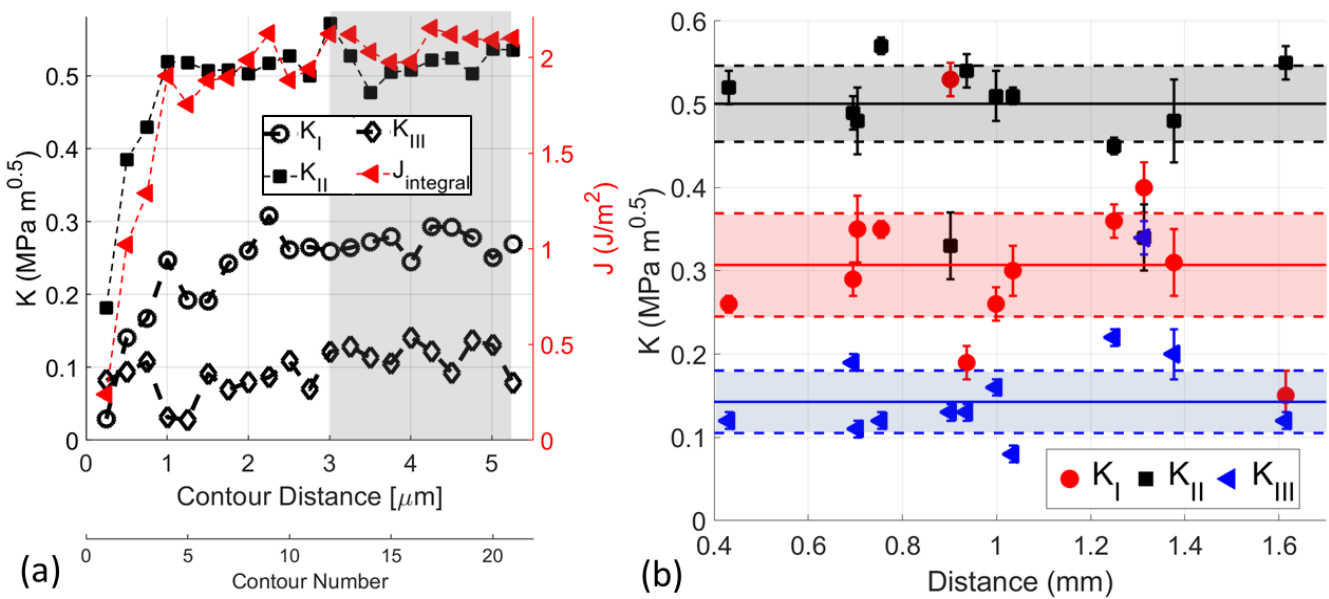

*Figure 5: (a) $J$-integral and three-dimensional stress intensity factors as the contour of the integration domain expanded from the crack tip. The example is at the crack tip position labelled 1 in Figure 3a. Values of the mode-specific $J$-integral (or equivalent stress intensity factor linked through the effective Youngs' modulus, shear modulus and Poisson's ratio estimated from the (001) silicon anisotropic stiffness tensor) are the average of the 10 values in the converged shaded region. The total $J$-integral is the summation of the mode-specific $J$-integrals. The values for all observations of this crack are given in Table 1. (b) Stress intensity factors through the observed intervals of the crack propagation. The horizontal axis is the total distance of crack propagation from initiation. The statistical bounds were obtained by the Bisquare weights method, implemented natively in MATLAB®, which uses an iterative reweighted least-square approach that minimises outliers' effect to calculate the mean and lower and upper bounds (shown above) or variant ($s^2$ in Figure 6).*

The crack plane and propagation direction are effectively constant between locations 5 and 12 (Figure 3), as the crack propagated on the (131) plane in the [$\bar{3}$10] direction with an average total energy release rate ($J$-integral) of 2.23 ± 0.17 J m$^{-2}$. The average mode I stress intensity factor is 0.31 ± 0.11 MPa m$^{0.5}$ with an average mode II stress intensity factor of 0.48 ± 0.06 MPa m$^{0.5}$. The average mode III stress intensity factor is small at 0.15 ± 0.06 MPa m$^{0.5}$. The average effective stress intensity factor, $K_{\text{eff}}$, for the (131) plane is 0.61 ± 0.02 MPa m$^{0.5}$. All means and variances are calculated using the bi-square weights method. The values for all 12 observations are reported in Table 1.

*Table 1: Geometrical information for the example crack (Figure 3) propagating in a* (001) *silicon wafer (angles are defined in Figure 7). The elastic moduli are evaluated normal to crack plane. The evaluated J-Integral, three-dimensional stress intensity factors ($K_I$, $K_{II}$, $K_{III}$) and effective stress intensity factor ($K_{eff}$) are also shown.*

| Label | Nominal applied compressive stress (MPa) | Increment of crack growth (µm) | Crack Trace Angles (°) $\emptyset, \theta$ | Trace | $E$ (GPa) | $\mu$ (GPa) | $v$ | $J$ (J m$^{-2}$) | $K_{\mathrm{eff}}$ (MPa m$^{0.5}$) | $K_I$ (MPa m$^{0.5}$) | $K_{II}$ (MPa m$^{0.5}$) | $K_{III}$ (MPa m$^{0.5}$) |
|---|---|---|---|---|---|---|---|---|---|---|---|---|
| 1 | 2.6 | 0 | 44.6, 20.6 | $[\bar{1}00](031) \pm 2.1°$ | 177.2 | 79.6 | 0.113 | 2.03 ± 0.10 | 0.60 ± 0.02 | 0.26 ± 0.01 | 0.52 ± 0.02 | 0.12 ± 0.01 |
| 2 | 11.8 | 263.3 | 40.1, 18.7 | $[\bar{1}00](031) \pm 0.2°$ | 176.8 | 79.2 | 0.116 | 2.08 ± 0.15 | 0.60 ± 0.02 | 0.29 ± 0.02 | 0.49 ± 0.02 | 0.19 ± 0.01 |
| 3 | 11.9 | 9.3 | 36.8, 19.0 | $[\bar{7}10](173) \pm 3.9°$ | 176.0 | 78.4 | 0.122 | 2.26 ± 0.26 | 0.62 ± 0.04 | 0.35 ± 0.04 | 0.48 ± 0.04 | 0.11 ± 0.01 |
| 4 | 13.3 | 50.9 | 30.8, 23.7 | $[\bar{4}10](142) \pm 2.2°$ | 173.9 | 76.4 | 0.138 | 2.71 ± 0.09 | 0.69 ± 0.01 | 0.35 ± 0.01 | 0.57 ± 0.01 | 0.12 ± 0.01 |
| 5 | 20.7 | 147.4 | 27.3, 23.3 | $[\bar{3}10](131) \pm 5.8°$ | 172.3 | 74.9 | 0.15 | 2.48 ± 0.19 | 0.65 ± 0.03 | 0.53 ± 0.02 | 0.33 ± 0.04 | 0.13 ± 0.01 |
| 6 | 21.1 | 34.6 | 26.4, 25.5 | $[\bar{3}10](131) \pm 8.0°$ | 171.8 | 74.5 | 0.154 | 2.07 ± 0.13 | 0.59 ± 0.02 | 0.19 ± 0.02 | 0.54 ± 0.02 | 0.13 ± 0.01 |
| 7 | 22.6 | 61.8 | 23.6, 24.7 | $[\bar{3}10](131) \pm 7.1°$ | 170.4 | 73.2 | 0.164 | 2.15 ± 0.16 | 0.60 ± 0.02 | 0.26 ± 0.02 | 0.51 ± 0.03 | 0.16 ± 0.01 |
| 8 | 23.2 | 36.5 | 24.9, 23.7 | $[\bar{3}10](131) \pm 6.1°$ | 171.1 | 73.8 | 0.159 | 2.12 ± 0.14 | 0.60 ± 0.02 | 0.3 ± 0.03 | 0.51 ± 0.01 | 0.08 ± 0.01 |
| 9 | 25.8 | 215.0 | 25.1, 20.6 | $[\bar{3}10](131) \pm 3.1°$ | 171.2 | 73.9 | 0.158 | 2.28 ± 0.14 | 0.62 ± 0.02 | 0.36 ± 0.02 | 0.45 ± 0.01 | 0.22 ± 0.01 |
| 10 | 26.5 | 63.5 | 25.3, 19.6 | $[\bar{3}10](131) \pm 2.1°$ | 171.3 | 74.0 | 0.158 | 2.52 ± 0.16 | 0.65 ± 0.02 | 0.40 ± 0.03 | 0.34 ± 0.04 | 0.34 ± 0.02 |
| 11 | 27.1 | 62.9 | 26.4. 17.9 | $[\bar{3}10](131) \pm 0.4°$ | 171.8 | 74.5 | 0.154 | 2.39 ± 0.34 | 0.62 ± 0.05 | 0.31 ± 0.04 | 0.48 ± 0.05 | 0.20 ± 0.03 |
| 12 | 30.7 | 237.8 | 25.5, 14.9 | $[\bar{3}10](131) \pm 2.6°$ | 171.4 | 74.1 | 0.157 | 2.04 ± 0.16 | 0.59 ± 0.02 | 0.15 ± 0.03 | 0.55 ± 0.02 | 0.12 ± 0.01 |

## 4. Discussion

Single crystal silicon is a brittle material at room temperature that can be most easily cleaved on the {110}, {111} and {001} planes (Table 2) [51]. Cleavage favourably propagates in the <110> direction [51], and experimentally measured fracture toughness varies both with the crack plane and direction [52,53]. Under a complex stress state, cleavage has been observed on other low-index planes of the {11/} type [54] and can also deviate from the <110> direction [55]. The geometrically necessary dislocation (GND) density at the crack tip (Figure 4a), which represents significant lattice rotations in only a small zone, demonstrates that the fracture was brittle with no significant plastic deformation, unlike cleavage in single-crystal metals [56,57]. The significant observations were obtained for a crack that propagated close to a low index {113} plane in a quasi-static manner with no branching (Table 1 and Figure 3). The initially complex crack path, which was not analysed, may be due to multi-axial stresses from the Hertzian contact of the rough edge of the sample with the loading anvil.

*Table 2: Recommended and reported mode I fracture toughness values for single crystal silicon cleavage at room temperature [51].*

| Fracture Plane | {100} | {110} | {111} |
|---|---|---|---|
| Reported $K_{IC}$ (MPa m$^{1/2}$) | 0.75–1.29 | 0.68–1.19 | 0.62–1.22 |
| Recommended $K_{IC}$ (MPa m$^{1/2}$) | 0.75 | 0.71 | 0.62 |
| Recommended J$_{IC}$ (J m$^{-2}$) | 2.163 | 1.483 | 1.022 |

The three-dimensional stress intensity factors (Table 1, Figure 5b) obtained for the crack show that the stress field surrounding the crack tip was almost constant during quasi-static cleavage along the (131) plane. The mixed-mode I/II loading was dominated by mode II, though some notable changes to higher mode I and lower mode II occurred at locations 5 and 10. The mode III stress intensity factor was generally small. The local variations in the relative modes may be due to slight length-scale deviations in the direction of the crack when it arrests. The observed 'Wallner' lines are evidence of crack deviation at arrest and re-initiation. Incorrect definition of the local crack propagation direction ($q_1$), which is assumed in the analysis to be the visible crack trace, would influence the symmetry of the field decomposition and, consequently, the inferred SIFs. Still, it would not significantly alter the effective stress intensity factor (see supplementary information) [58].

The average mode I stress intensity factor measured for cleavage on (131) is lower than the reported mode I fracture toughness values of the low-index cleavage planes (Table 2); however, this is to be expected. There are various criteria proposed for crack propagation under mixed-mode loading, of which the maximum potential energy release rate (MPERR), developed for both 2D [59–61] and 3D [62], is robust and simple to apply. Chang *et al.* [62] used a three-dimensional fracture surface to represent the mixed-mode contributions to the MPERR criterion (Figure 6a and c). The experimental SIFs (Table 1) for (131) cleavage (best) fit the critical surface for $K_{IC}$ of 0.69 MPa m$^{0.5}$ with R$^2$ equals 0.84 (Supplementary information). Thus, considering the quality of the fit (i.e., uncertainty in $K_{IC}$) $K_{IC}$ is 0.69 ± 0.03 MPa m$^{0.5}$.

$$\left(K_I / K_{IC}\right)^2 + \left(K_{II} / K_{IIC}\right)^2 + \left(K_{III} / K_{IIIC}\right)^2 = 1 \tag{12}$$

$$K_{IIC} = \frac{\sqrt{3}}{2} K_{IC}, \qquad K_{IIIC} = \frac{\sqrt{k+1}}{2} K_{IC} \tag{13}$$

$$k = \frac{(3-v)}{(1+v)} \text{ for plane stress} \tag{14}$$

$$\psi = \tan^{-1}\left(K_{III} / \sqrt{K_I^2 + K_{II}^2}\right) \tag{15}$$

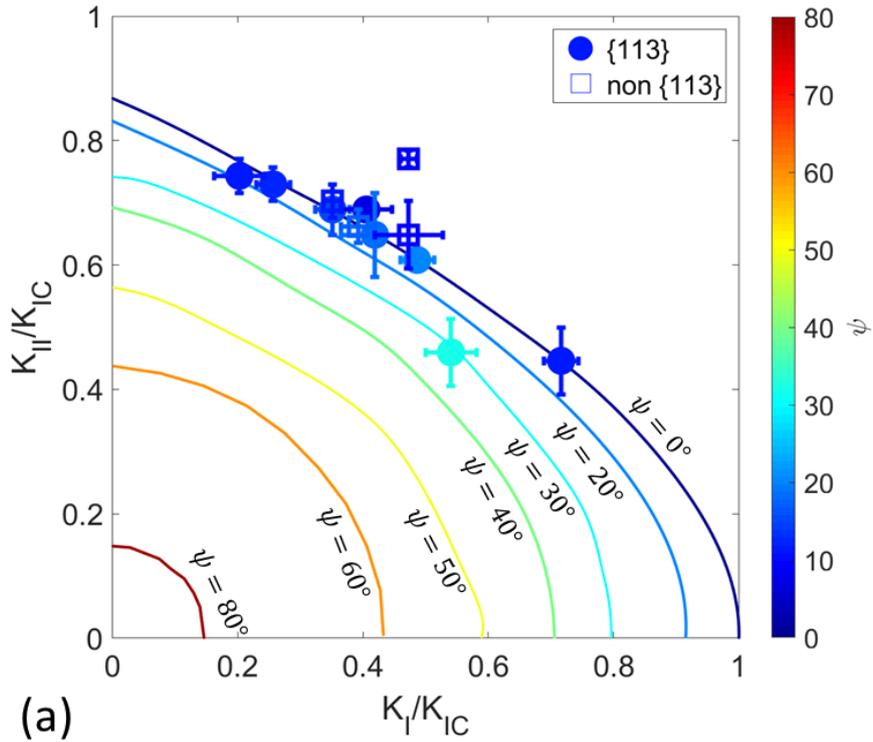

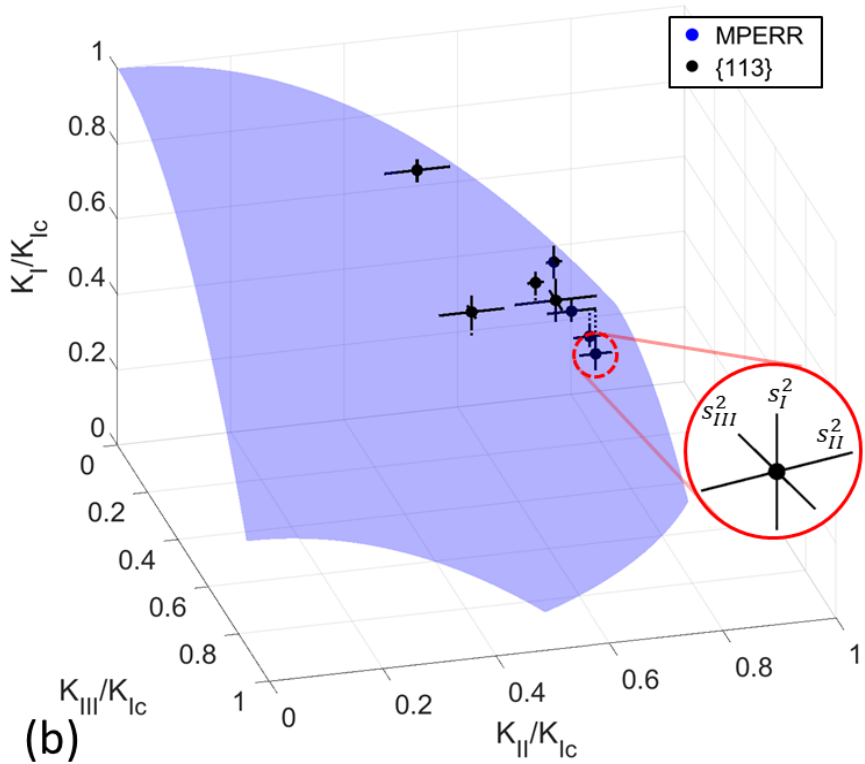

Figure 6: Experimental data (Table 1) with the critical surface (equations (12) to (15)) constructed at a constant value of $K_{IC}$ = 0.69 MPa m$^{0.5}$ ($R^2$ = 0.84) for the mixed-mode maximum potential energy release rate (MPERR) criterion [62]. (a) All data (1 to 12) on a contour plot with $K_{IC}$ = 0.69 MPa m$^{0.5}$ (see Supplementary information). (b) three-dimensional plot of the critical to the experimental data for (131) (i.e., points 5 to 12). The solid black lines are the measurement uncertainty. The vertical dotted black lines indicate the difference (parallel to the $K_I/K_{IC}$ axis) between the data and the surface.

This novel analysis method provides a direct and high-resolution solution to quantify the elastic fields acting on a crack tip, using local measurements without knowing the external boundary conditions (e.g., load, crack length). The analysis can be applied to mixed mode loading and the in situ studies of quasi-static cleavage crack propagation. The present analysis is suitable for elastic deformation or small-scale yielding conditions, but more advanced decomposition algorithms will need to be employed for significant elasto-plastic deformation [63]. This local method has advantages over micro-mechanical methods that use geometrical analytical solutions, as these rely on accurate knowledge or control of the external boundary conditions such as specimen dimensions, load, and displacement.

The present study employed a brittle single silicon crystal where the deformation is localised at the crack tip and with careful consideration of the reference pattern selection [38]; thus, reducing the significant problem of obtaining a stress-free HR-EBSD reference pattern that typically arises in the analysis of engineering materials [64]. This issue needs to be addressed before the method can be more generally applied. Although that remains challenging, this might be achieved by simulation of reference patterns [65]. Other possible experimental approaches are high-resolution strain mapping methods that provide absolute strains, such as cross-correlation of transmission electron microscopy (TEM) acquired nano-diffraction patterns [66], Laue microdiffraction [67], and dark-field X-ray microscopy [68]. As this method can be used to examine mixed-mode fracture criteria with a high spatial resolution, it would allow investigations of the fracture properties of brittle inclusions, coatings, interfaces, and grain boundaries in engineering materials.

# 5. Conclusion

A novel approach has been formulated to decompose the $J$-integral evaluation of the elastic energy release rate to the three-dimensional stress intensity factors ($K_{I,II,III}$) directly from experimental measurements of the local elastic deformation gradient tensors of the crack field under small-scale yielding conditions. Strictly speaking, no numerical method, such as Finite Element or Finite Difference, has been used to solve the boundary value problem.

An exemplar study is presented of a quasi-static crack, inclined to the observed surface, propagating on low index $\{hkl\}$ planes in a (001) single crystal silicon wafer and was characterised using in situ high (angular) resolution electron backscatter diffraction (HR-EBSD). The mixed-mode crack fields were consistent with a constant maximum potential energy release rate criterion for crack propagation, with average fracture toughness, $K_{IC}$, of the {113} plane measured to be 0.69 ± 0.03 MPa m$^{0.5}$. This high-resolution approach can potentially be used to study the local fracture resistance of brittle inclusions, coatings, and interfaces at the micron scale within material microstructures under complex loading.

# Acknowledgements

We are grateful to Professor Peter Wilshaw (University of Oxford) for supplying the specimen material. The authors acknowledge the use of experimental equipment from the Laboratory for In situ Microscopy and Analysis (LIMA) and characterisation facilities within the David Cockayne Centre for Electron Microscopy (DCCEM), Department of Materials, University of Oxford, alongside financial support provided by the Henry Royce Institute (Grant ref EP/R010145/1). Abdalrhaman Koko is grateful to Engineering and Physical Sciences Research Council (EPSRC) for providing PhD studentship (Grant ref EP/N509711/1).

# Authorship Contribution Statement

**Abdalrhaman M. Koko:** Conceptualisation, Methodology, Software, Investigation, Formal analysis, Writing - original draft, Visualisation.

**Thorsten H. Becker:** Methodology, Software, Writing - review & editing.

**Elsiddig Elmukashfi:** Formal analysis.

**Nicola M. Pugno:** Writing - review & editing.

**Angus J. Wilkinson**: Software.

**T. James Marrow:** Conceptualisation, Resources, Writing - review & editing, Supervision, Funding Acquisition.

# Data and code availability

The relevant codes and data to reproduce the results presented in this paper are available at https://doi.org/10.5281/zenodo.6411454. The MATLAB® implementation of the method and a benchmarking dataset are available at https://doi.org/10.5281/zenodo.6411485.

# HR-EBSD analysis of in situ stable crack growth at the micron scale - Supplementary information

A. Koko *et al.*

## Contents

## Supplementary Figures

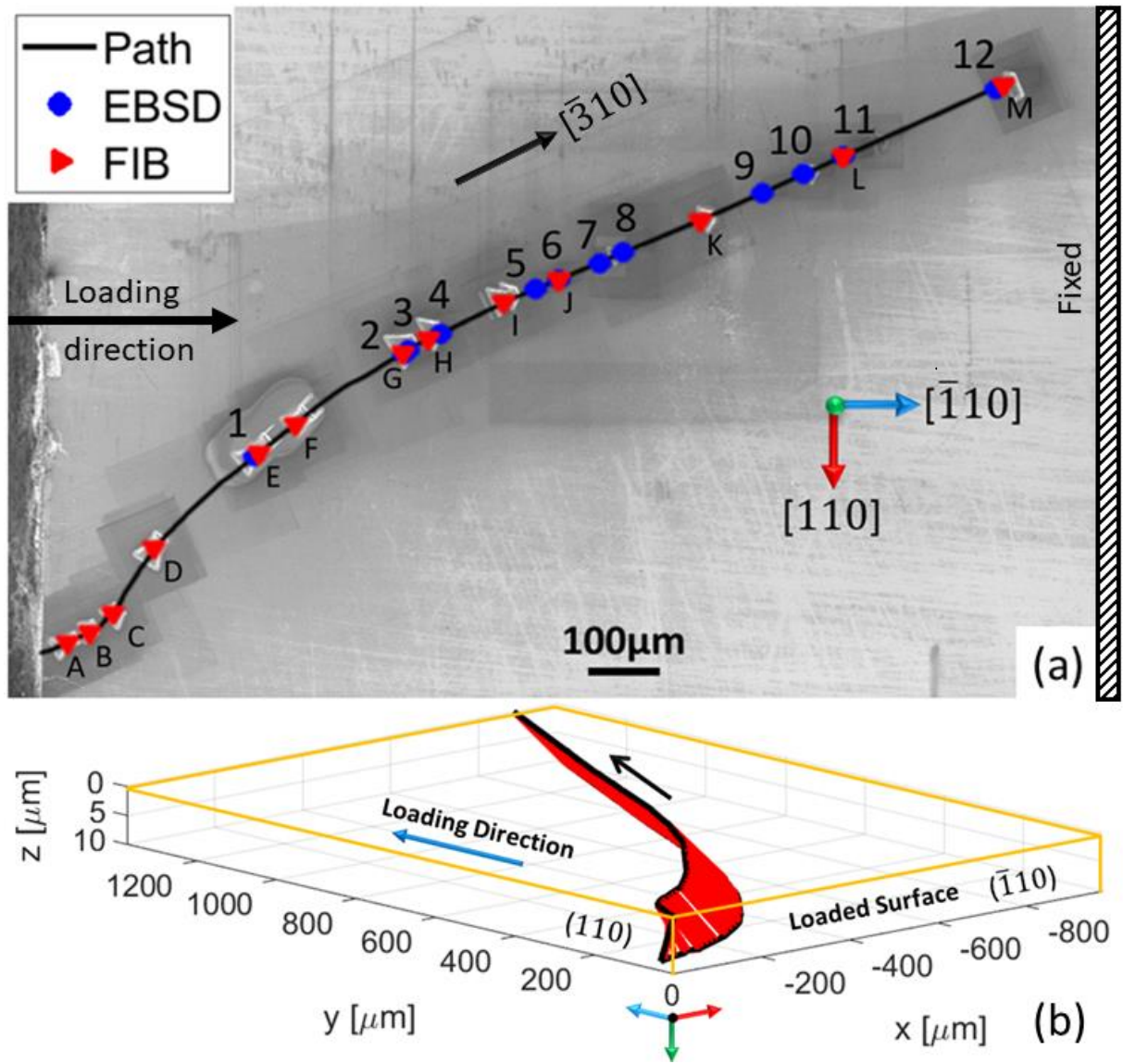

*Supplementary Figure 1: The exemplar crack propagated in a (001) silicon wafer, loaded in compression parallel to $[\bar{1}10]$: a) showing the crack path and HR-EBSD mapping and FIB cutting locations. The numbers mark the intervals of crack growth, after which the stable crack tip was observed in situ under load by HR-EBSD. The sample surface was cleaned from the accumulated debris before FIB cutting of trenches parallel to (110) at the marked locations to examine the sub-surface crack geometry; b) 3D visualisation of the crack geometry close to the surface, using the FIB slice traces and cubic interpolation between points.*

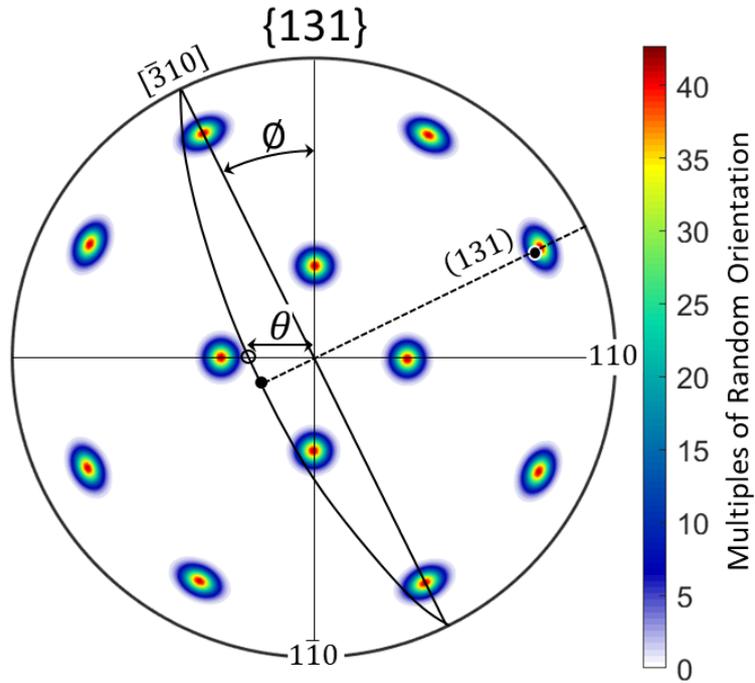

*Supplementary Figure 2: EBSD data confirmed the (001) wafer orientation and defined the x and y axes as [110] and [1̄10] as shown in Figure 3 in the main text. The pole figure and measured traces were then used to identify the crack direction and plane. The example here is for observation 11, showing the crack pole is close to a {131} plane.*

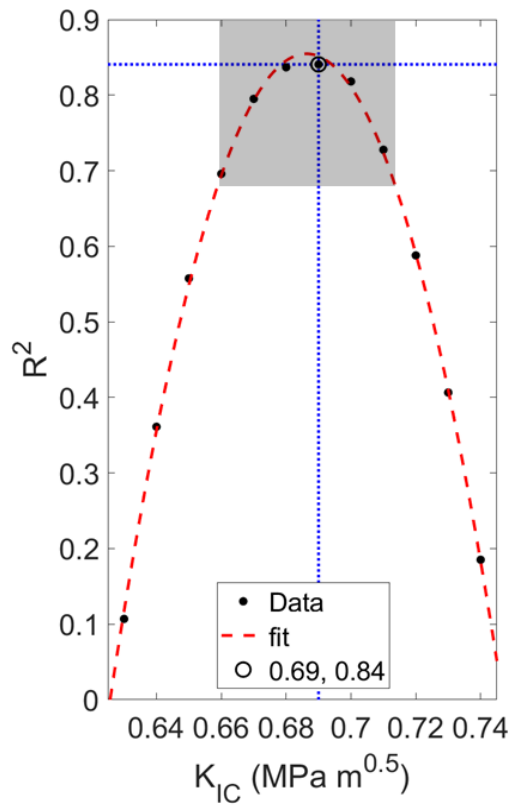

*Supplementary Figure 3: The goodness of the fit ($R^2$) using different $K_{IC}$ values [1] on (131) plane data with the grey area highlighting weighted uncertainty.*

# Supplementary Notes (A): Assumed Crack Direction

There is no significant effect of the beam on the sample. Bouscaud *et al.* [2] considered how the electron beam could cause the local temperature to increase in the interaction volume due to dissipating energy created from inelastic scattering. The interaction volume can be estimated using Monte Carlo Simulation (Casino v2.48 [3]). A 173 nm effective volume of interaction (i.e., depth of resolution, $r$) was calculated for 5 x $10^6$ electrons bombarding a 70° titled silicon sample placed inside a Merlin FEG-SEM that has a beam mean-radius of ~25nm at 20 keV beam energy ($E_0$) of and 10 nm current ($I_0$) (Supplementary Figure 4). A (mean) induced temperature ($\Delta T$) of 1.2 K was then estimated from equation (1), assuming a thermal conductivity ($\lambda$) of 149 W m$^{-1}$ K$^{-1}$ at 300 K [4], which is negligible.

$$\Delta T = \frac{E_0 I_0}{2\pi\lambda r} \tag{1}$$

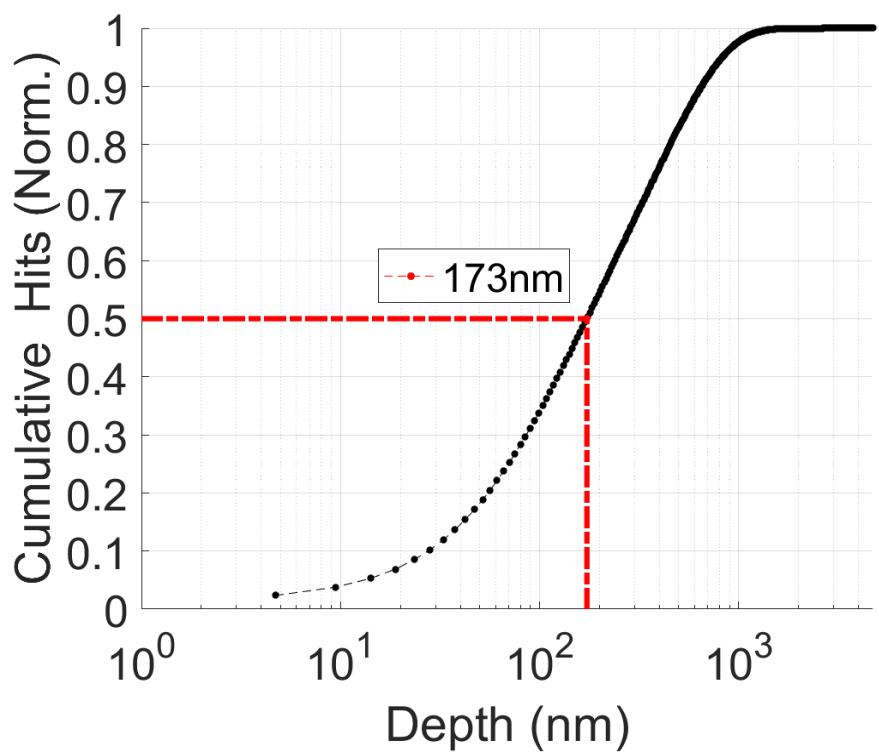

*Supplementary Figure 4: Cumulative probability of backscattered electron detection vs depth from the surface for Monte Carlo simulated electron trajectories in silicon. A dashed red line indicates the mean of the distribution.*

The fracture surface was marked by 'Wallner' lines that indicate local deviations in the crack plane. Examples of the surface observations (Supplementary Figure 5c) show that the arrested crack is sometimes slightly kinked and returns to the previous plane on propagation. The kinks may be due to the effects of local crack tip plastic deformation as the crack arrests (see GND density map in Figure 4a in the main text). The slight changes in local crack direction mean that the assumed crack (expected) propagation direction ($q$) – which was assumed to follow the average crack plane geometry – is not always applicable. Such an error will influence the mode mixity (i.e., the relative magnitude of SIFs) but not the effective stress intensity factor ($K_{\text{eff}}$) because the loading modes are coupled [5,6].

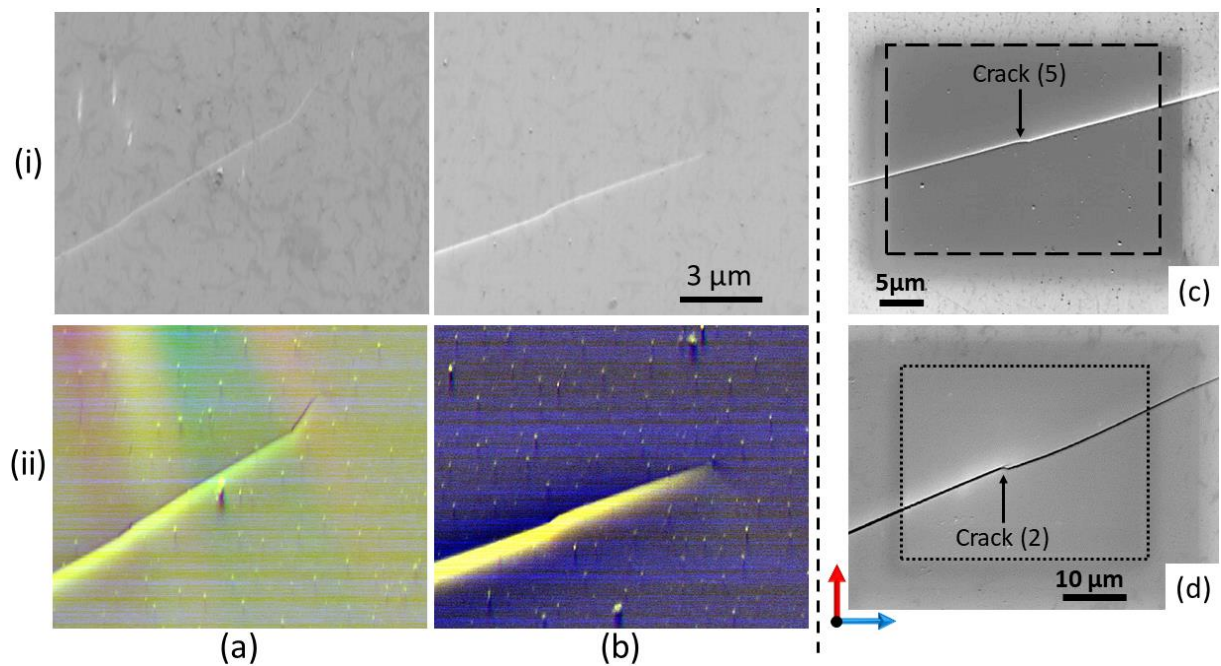

*Supplementary Figure 5: (i) Secondary electron, and (ii) forward scatter detector (FSD) images for the crack tip at the points labelled (a) 1 and (b) 4 in Supplementary Figure 1. (c) and (d) are the crack trace after propagation at the following in-situ observation.*

This is illustrated in Supplementary Figure 6 for a case where a mode I field has been simulated in the ABAQUS® Finite element solver, and mode I and II stress intensity factors have been calculated with the assumption of $q$-vectors that deviate from the actual direction using the interaction integral method [7] natively implemented in ABAQUS®. As the assumed direction deviates from the true direction with the maximum potential strain energy release rate ($J$-integral), the mode II stress intensity factor (SIF) increases, and the mode I SIF decreases.

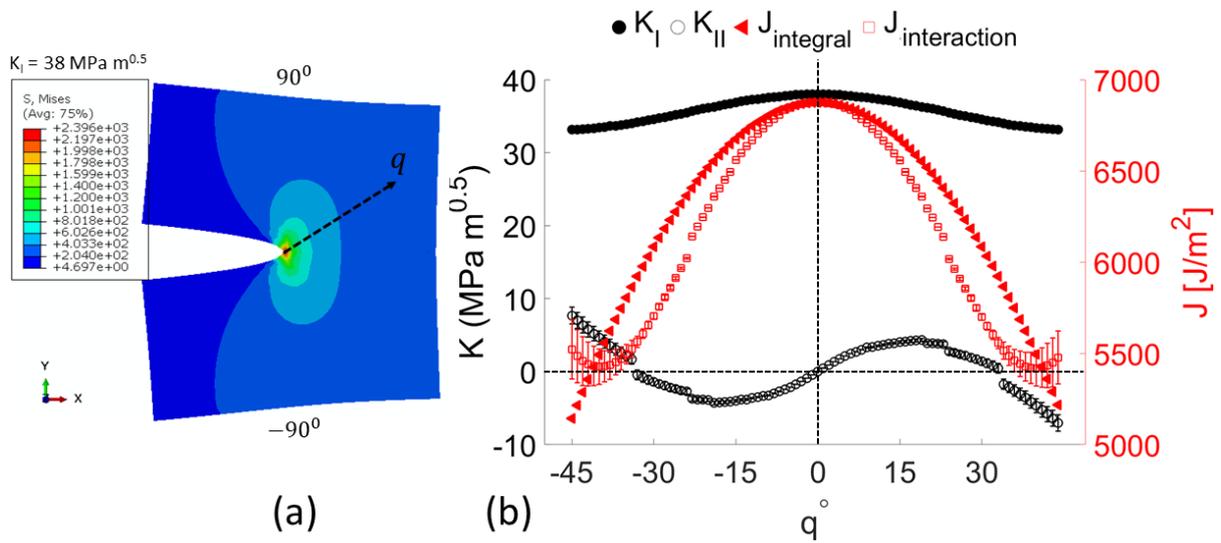

*Supplementary Figure 6: (a) ABAQUS® simulation of a two-dimensional crack on an isotropic material (E = 210 GPa), loaded to 38 MPa m^{0.5} in mode I. (b) Calculated mode I and II while changing the assumed crack extension direction (q) using the interaction integral method (natively implemented in ABAQUS®). As the crack, it experiences increased mode II and reduced mode I.*

## Supplementary Notes (B): In situ EBSP$_0$ selection

The effect of using a 'Native' compared to 'Chosen'[1] on the correlation parameters was quantified and tabulated in Table 1. PH and MAE values are more consistent between maps that used the 'Chosen' approach than those using the 'Native' approach. The accuracy of the maps was slightly improved using the 'Chosen' EBSP$_0$, which was then used for further *J*-integral analysis.

*Table 1: Effect of using 'Native', 'Chosen', and 8$^{th}$ interval EBSP$_0$ as a reference pattern.*

| Interval | Native | | Chosen | | 8$^{th}$ Chosen-EBSP$_0$ | |
|---|---|---|---|---|---|---|
| | PH | MAE ($10^{-4}$) | PH | MAE ($10^{-4}$) | PH | MAE ($10^{-4}$) |
| 1 | 0.75 | 1.8 | 0.74 | 1.6 | 0.75 | 17.9 |
| 2 | 0.75 | 1.8 | 0.74 | 1.6 | 0.73 | 7.9 |
| 3 | 0.69 | 1.9 | 0.71 | 1.7 | 0.74 | 5.1 |
| 4 | 0.74 | 1.9 | 0.74 | 1.6 | 0.74 | 14.8 |
| 5 | 0.74 | 1.8 | 0.74 | 1.6 | 0.74 | 5.3 |
| 6 | 0.74 | 1.8 | 0.73 | 1.6 | 0.74 | 5.0 |
| 7 | 0.74 | 2.1 | 0.73 | 1.7 | 0.78 | 2.3 |
| 8 | 0.73 | 1.8 | 0.73 | 1.6 | 0.73 | 1.6 |
| 9 | 0.74 | 1.9 | 0.74 | 1.7 | 0.74 | 10.8 |
| 10 | 0.74 | 1.8 | 0.73 | 1.6 | 0.76 | 18.7 |
| 11 | 0.75 | 1.9 | 0.75 | 1.6 | 0.76 | 21.1 |
| 12 | 0.74 | 1.7 | 0.74 | 1.6 | 0.74 | 19.6 |
| Average | 0.74 | 1.9 | 0.74 | 1.6 | 0.75 | 10.8 |

Since the beam conditions and experiment geometry were kept the same during the experiment intervals, all maps can be calculated relative to one EBSP$_0$ instead of 12. A synthetic 12 x 12 EBSD map was created using the $EBSP_0$ from each interval, as shown in the schematic in Supplementary Figure 1, where numbers donate the $EBSP_0$ interval. HR-EBSD analysis was conducted on the data. In the synthetic data, the 8th interval $EBSP_0$ was chosen as the least deformed from the available $EBSP_0$ using conventional selection criteria, i.e., based on pattern quality (IQ) and low kernel average misorientation (KAM).

---

[1] 'Native' and 'Chosen' EBSP$_0$ definition can be found in Ref. [27]. Briefly, 'Native' is EBSP$_0$ selected away from stress concentration as identified using pattern image quality (IQ) and the grain mean orientation, where 'Chosen' it is the point with the give least mean angular error and highest cross-correlation peak height.

| 1 | 2 | 3 | 4 | 5 | 6 | 7 | 8 | 9 | 10 | 11 | 12 |
|---|---|---|---|---|---|---|---|---|----|----|----|
| 1 | 2 | 3 | 4 | 5 | 6 | 7 | 8 | 9 | 10 | 11 | 12 |
| 1 | 2 | 3 | 4 | 5 | 6 | 7 | 8 | 9 | 10 | 11 | 12 |
| 1 | 2 | 3 | 4 | 5 | 6 | 7 | 8 | 9 | 10 | 11 | 12 |
| 1 | 2 | 3 | 4 | 5 | 6 | 7 | 8 | 9 | 10 | 11 | 12 |
| 1 | 2 | 3 | 4 | 5 | 6 | 7 | 8 | 9 | 10 | 11 | 12 |
| 1 | 2 | 3 | 4 | 5 | 6 | 7 | 8 | 9 | 10 | 11 | 12 |
| 1 | 2 | 3 | 4 | 5 | 6 | 7 | 8 | 9 | 10 | 11 | 12 |
| 1 | 2 | 3 | 4 | 5 | 6 | 7 | 8 | 9 | 10 | 11 | 12 |
| 1 | 2 | 3 | 4 | 5 | 6 | 7 | 8 | 9 | 10 | 11 | 12 |
| 1 | 2 | 3 | 4 | 5 | 6 | 7 | 8 | 9 | 10 | 11 | 12 |
| 1 | 2 | 3 | 4 | 5 | 6 | 7 | 8 | 9 | 10 | 11 | 12 |

*Supplementary Figure 1: Schematic of patterns arrangement in the synthetic EBSD map.*

Note that a shadow from the stage anvil was cast on the patterns on the first interval (Supplementary Figure 02a) due to the proximity of the crack to the anvil; however, once excluded, it did not affect HR-EBSD cross-correlation (see Table 1). From HR-EBSD analysis (Supplementary Figure 02b), the 8th and 7th intervals show a relatively low deformation, but deformation conditions exacerbate with intervals which can be illustrated clearly from the gradient seen in the in-plane components of the normal strain and $\omega_{13}$. The phantom (or artificial) stresses were introduced at each interval can be as high as 2.5 GPa and highly affect the in-plane stress (Supplementary Figure 02c).

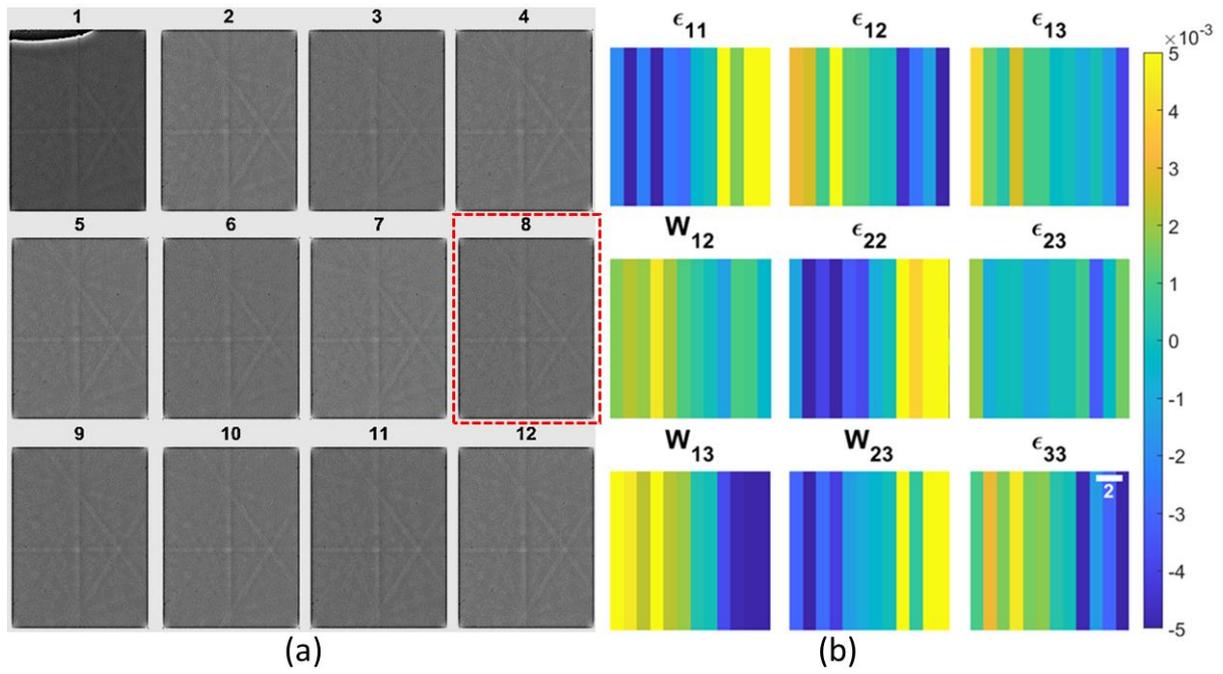

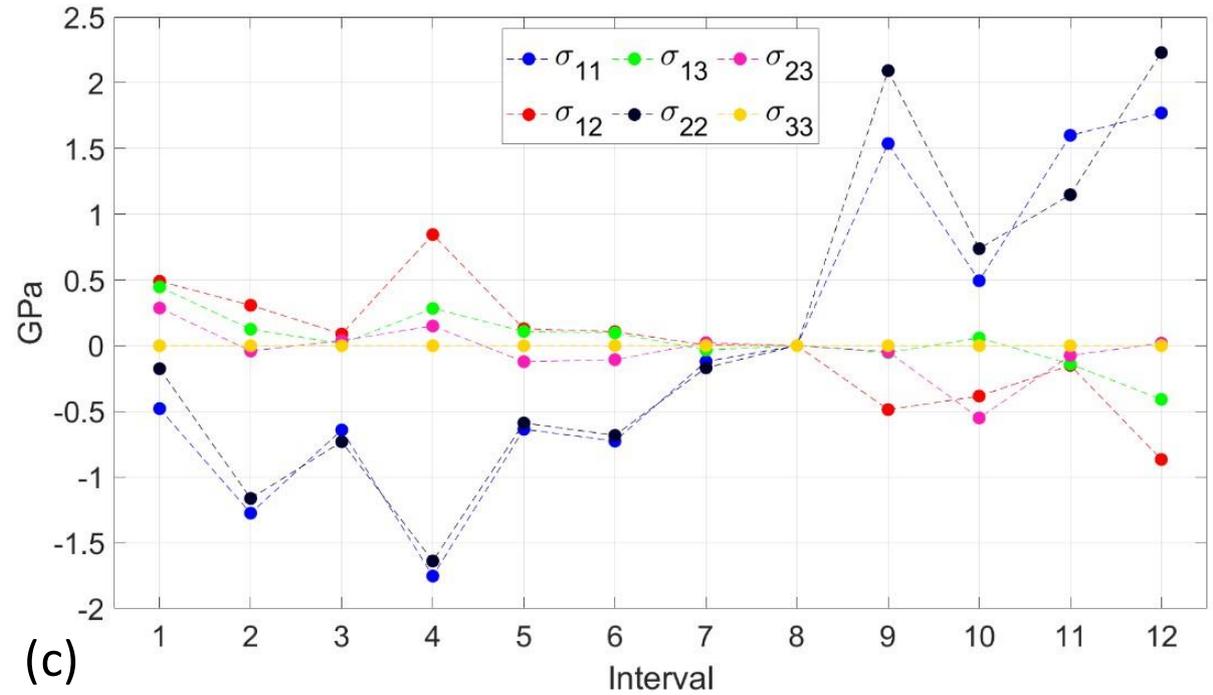

*Supplementary Figure 02: (a) In situ crack Chosen EBSP$_0$ at different experimental intervals. (b) Elastic deformation component from the Chosen EBSP$_0$ from each interval compared to EBSP$_0$ from the 8$^{th}$ interval (red squared in a). (c) Stress tensors relative to the 8$^{th}$ interval were calculated using (001) Silicon anisotropic elastic properties [8]. $\sigma_{33}$ is zero due to the assumed traction-free surface [9].*

A closer look can be taken at the effect of swapping EBSPs across scans using the 8$^{th}$ EBSP$_0$ for the 7$^{th}$ and 9$^{th}$ intervals. As indicated by the synthetic EBSD data in Supplementary Figure 02, the 7$^{th}$ interval deformation field slightly changes when using the 8$^{th}$ EBSP$_0$ as a reference pattern, as shown in Supplementary Figure 02c. As indicated in Supplementary Figure 02c, for the 9th scan, the deformation magnitude (in each tensor) was directly increased, especially in the in-plane components. This clearly shows, and contrary to our expectation, that each EBSD map has a different imaging condition; even when the current, voltage and working distance were kept constant, pattern centre (PC) shift due to beam movement during acquisition was minimised [10], pattern remapping [11], and no further angular effect from drift was expected as the ROI was small (~20 x 15 $\mu m^2$) and the sample was conductive [12].

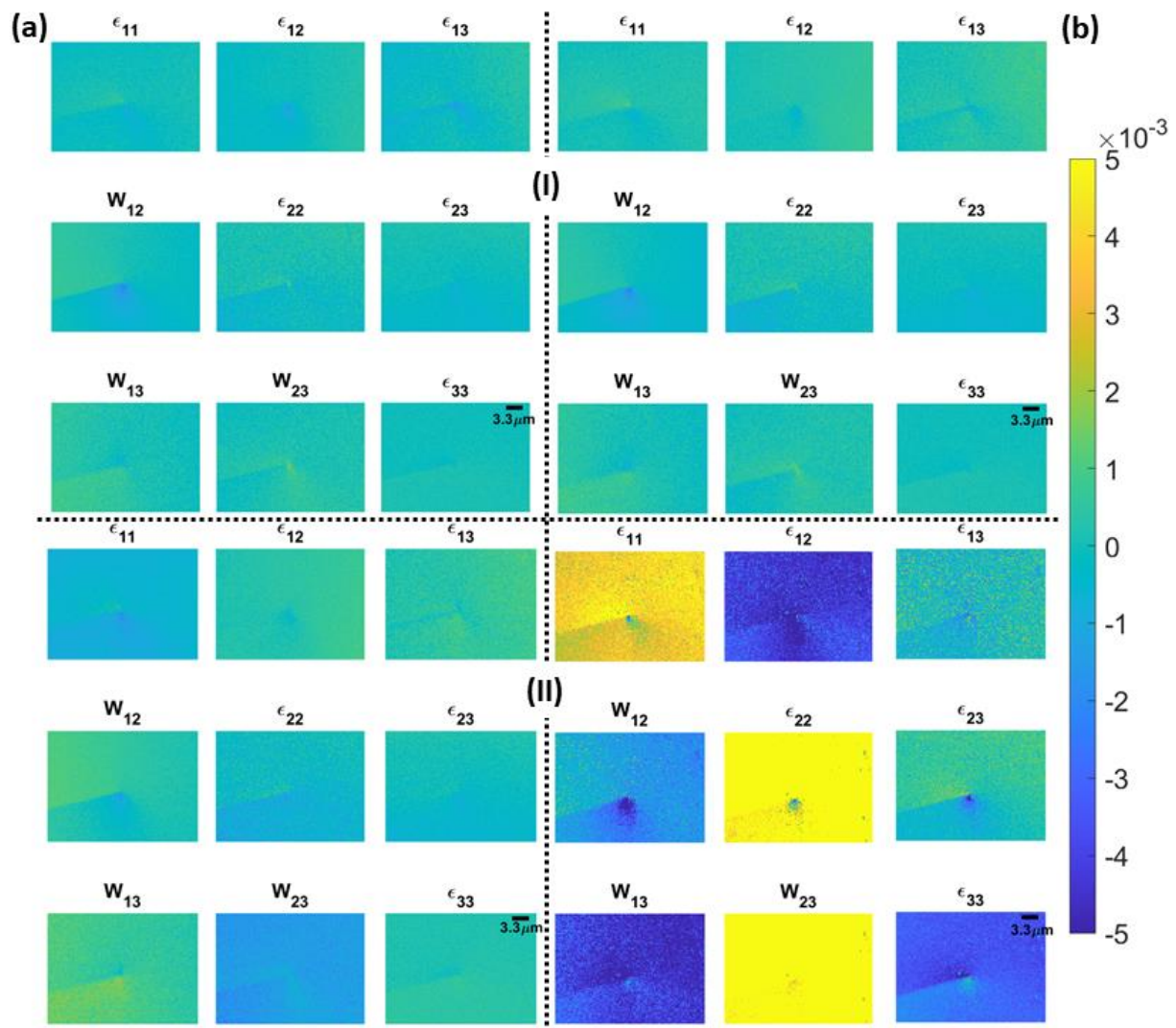

*Supplementary Figure 3: Deformation field components for steps (a) 7 and (b) 9 of in situ test using (I) Chosen and (II) Swapped (from step 8) EBSP$_0$.*

It appears that slight stage movement coupled with minor re-focusing of the field of view (although there is no necking or change of thickness effect here) changed the experiment geometry; thus, impacting the probed EBSPs. This shows that swapping EBSP$_0$ across intervals or finding an EBSP$_0$ for all intervals and relating all 'relative' (or absolute) information to one EBSP$_0$ will yield erroneous results. However, it slightly improves the correlation (see Table 1).

The results are similar to results used pristine single crystal mounted within the sample for calibration [13] due to the change in projection parameters. When the 1$^{st}$ iteration of HR-EBSD calculates both real and artificial strain, the 2$^{nd}$ iteration will not separate the real strain (Supplementary Figure 3) due to pattern centre variation inducing phantom stress [14]. This directly contradicts Deal *et al*.'s [15] evidence of the feasibility of using a physical reference pattern (EBSP$_0$) library to calculate absolute strains using high (angular) resolution electron backscatter diffraction (HR-EBSD) with 11% repeatability. While the physical reference pattern (EBSP$_0$) library method might work for very limited cases [16], it seriously underestimates the error (or phantom stress) and yet faces the same limitations daunting simulated reference patterns for absolute strain measurement but without providing an understanding of the (physical) building blocks.

## Supplementary Notes (C): Analytical field benchmarking

A mixed-mode crack's displacement field that has a mode I stress intensity factor ($K_I$) of 3 MPa m$^{0.5}$, mode II ($K_{II}$) of 1 MPa m$^{0.5}$, and mode III ($K_{III}$) of 2 MPa m$^{0.5}$ were created using an analytical solution [17] and assuming plane stress conditions (eq. 1 to 5). The elastic modulus ($E$) and Poison's ratio ($v$) were 210 GPa and 0.3, respectively. The field of view was 1*1m, having 0.02*0.02 square elements and a crack tip at the centre (0,0). Then, the (numerical) displacement gradient ($u_{i,j}$) calculated from the displacement (equation 2-7).

$$u_1 = \frac{K_I}{2\mu}\sqrt{\frac{r}{2\pi}}\cos\left(\frac{\theta}{2}\right)\left[k - 1 + 2\sin^2\left(\frac{\theta}{2}\right)\right] + \frac{K_{II}}{2\mu}\sqrt{\frac{r}{2\pi}}\sin\left(\frac{\theta}{2}\right)\left[k + 1 + 2\cos^2\left(\frac{\theta}{2}\right)\right] \qquad 1$$

$$u_2 = \frac{K_I}{2\mu}\sqrt{\frac{r}{2\pi}}\cos\left(\frac{\theta}{2}\right)\left[k + 1 - 2\cos^2\left(\frac{\theta}{2}\right)\right] - \frac{K_{II}}{2\mu}\sqrt{\frac{r}{2\pi}}\cos\left(\frac{\theta}{2}\right)\left[k - 1 - 2\sin^2\left(\frac{\theta}{2}\right)\right] \qquad 2$$

$$u_3 = \frac{2K_{III}}{\mu}\sqrt{\frac{r}{2\pi}}\sin\left(\frac{\theta}{2}\right) \qquad 3$$

$$\text{shear modulus } (\mu) = \frac{E}{2(1+v)} \qquad 4$$

$$k = \frac{3-v}{1+v} \qquad 5$$

$$u_{i,j} = \frac{\partial u_i}{\partial x_j} \qquad 6$$

Equation (7) was then used to decompose the displacement gradient to different modes. The (infinitesimal) strain was then calculated from each mode with their relationship as expressed below and shown in Supplementary Figure 1a.

$$\varepsilon_{ij} = \frac{1}{2}\left(u_{i,j} + u_{j,i}\right) \qquad 0-7$$

Using the displacement derivatives with the method outlined in the main text, the calculated *J*-integral and decomposed stress intensity factors matched inputted values (Supplementary Figure 1b). Highly localised fields close to crack edges can also influence initial convergence, but convergence stabilises as the domain expands.

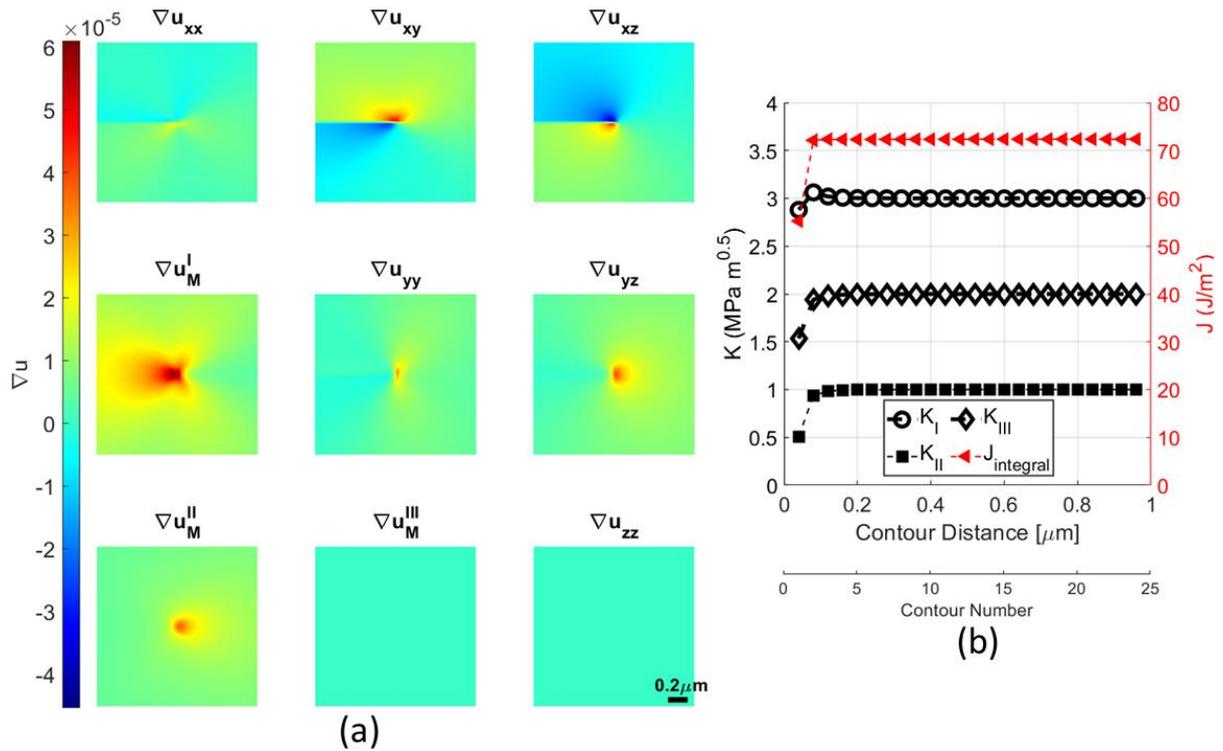

*Supplementary Figure 1: (a) The analytical displacement gradient for a stationary mixed loaded crack. (b) J-integral and decomposed loading modes as a function of contour distance from the crack tip.*

The key sources of errors in this analysis were assessed. First, a normally distributed random noise was incrementally induced from zero to 10% on all the strain components. Supplementary Figure 2a shows the amount of induced error in each component, with $K_{II}$ being highly influenced by noise. The dashed upper pound line indicates that this method is highly vulnerable to noise compared to a method that uses total elastic displacement, e.g., a 6% noise induces a convergence error of 7.6 ± 18.8 %, 16.8 ± 35.5% and 4.2 ± 11.2 % in $K_I$, $K_{II}$ and $K_{III}$, respectively, compared to 3.8%, 3% and 0.8% when using displacement field subjected to same noise [18,19]. This is because the induced noise on three displacement components will be diluted when the derivatives are calculated, where the induced noise on the nine strain components will directly affect the analysis. Calculating the strains next to the crack flanks is critical in mixed mode; this region cannot be masked or excluded, especially in mode I, but can be carefully extrapolated from the displacement gradients [20].

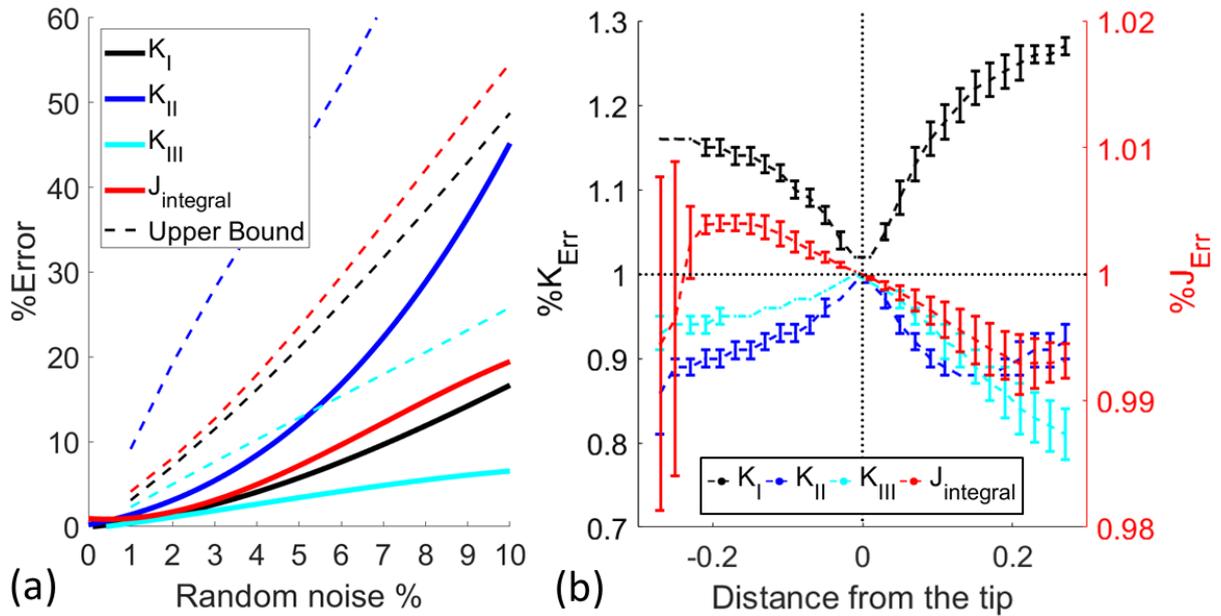

*Supplementary Figure 2: Error due to (a) noise and (b) crack tip accuracy.*

Second, the analysis sensitivity to crack tip location was assessed using the accurate crack tip positioned at the origin (0,0) coordinate. The error in accurately locating the crack increases if the location was assumed ahead relative to behind the crack (Supplementary Figure 2b). This is due to the nature of the analytical field, as there is symmetry on the crack edges, but the magnitude at the crack tip gradually changes with distance. Detailed analysis of two-dimensional errors reveals a distinct behaviour for each component of SIFs and the sum of all components in the J-integral influencing both magnitude and convergence.

Linear strain and rotation gradient, caused by pattern centre (PC) shift due to the beam movement during each EBSP acquisition, were not assessed as this spurious gradient magnitude and distribution are highly dependent on the $EBSP_0$ deformation status and position with respect to the beam. Nevertheless, errors due to PC shift can be reduced using AstroEBSD [10]. To less extent, certain meshing [21] and element/node configurations produce strain singularities [22,23] to improve accuracy; however, the discrepancy in results obtained from different singular elements persists [24,25]. For *J*-integral, comparably coarse meshes are sufficient, with no unique crack tip elements being required, which reduces the need for a high degree of mesh refinement at the crack tip [24].

In addition, this decomposition method is suitable for a regulated mesh and a crack field that is symmetric around the crack with no apparent noise at the crack edges but does not work for other stress raisers with compressive mechanical conditions (especially mode I) as minus values are interpreted as imaginary. Also, the stress field needs to be linearly elastic and have the $r^{-0.5}$ form of singularity, where $r$ is the distance from the crack tip. Complex and computationally expensive methods can be used as an alternative [26], which can be coupled with finite element solvers to get the most out of the measured field, including padding noise near the crack [20], calculate the crack probable extension direction (e.g., XFEM ABAQUS®), etc.